\documentclass[10pt]{article}
\usepackage{epsfig,graphicx,amssymb,amsfonts}

\def \be {\begin{equation}}
\def \ee {\end{equation}}
\def \bea {\begin{eqnarray}}
\def \eea {\end{eqnarray}}

\def \rr {\raise.35ex\hbox{\small $\prime$}\kern-.17em{\mbox{\large $\imath$}}}
\def \del {\partial}
\def \dels {\partial\kern-.5em / \kern.5em}
\def \As {{A\kern-.5em / \kern.5em}}
\def \Ds {D\kern-.7em / \kern.5em}

\def \a {\alpha}
\def \b {\beta}
\def \dag {\dagger}
\def \g {\gamma}

\def \d {\delta}
\def \eps {\epsilon}

\def \lam {\lambda}

\def \s {\sigma}

\def \om {\omega}

\def \th {\theta}

\def \t {\tau}

\def \II {I\hspace{-.1em}I\hspace{.1em}}

\def \IIB {\mbox{\II B\hspace{.2em}}}

\def \EOM {(\mbox{EOM})}
\def \D {{\cal D}}

\begin{document}
\begin{titlepage}

\begin{center}
\hfill hep-th/0401167\\
\vskip .5in

\textbf{\LARGE
Isometry of $AdS_2$ And The $c=1$ Matrix Model
}

\vskip .5in
{\large Pei-Ming Ho} 
\vskip 15pt

{\small Department of Physics,
Harvard University,
Cambridge, MA 02138, USA}
\footnote{On leave from National Taiwan University.}
\\
{\small Department of Physics,
National Taiwan University,
Taipei 106, Taiwan, R.O.C.}
\\
{\small Physics Division,
National Center for Theoretical Sciences,
National Taiwan University,
Taipei 106, Taiwan, R.O.C.}
\\

\vskip .2in
\sffamily{
pmho@phys.ntu.edu.tw}

\vspace{60pt}
\end{center}
\begin{abstract}

Implications of the $SL(2,\mathbb{R})$ symmetry of
the $c = 1$ matrix models are explored.
Based on the work of de Alfaro, Fubini and Furlan,
we note that when the Fermi sea is drained,
the matrix model for 2 dimensional string theory
in the linear dilaton background is equivalent to
the matrix model of $AdS_2$ recently proposed by Strominger,
for which $SL(2,\mathbb{R})$ is an isometry.
Utilizing its Lie algebra,
we find that a topological property of $AdS_2$
is responsible for quantizing D0-brane charges in type 0A theory.
We also show that the matrix model faithfully reflects
the relation between the Poincare patch and global coordinates of $AdS_2$.

\end{abstract}
\end{titlepage}
\setcounter{footnote}{0}

\section{Introduction}

Motivated by the problems with unstable D-branes,
the $c=1$ matrix model \cite{MM}
(for reviews see \cite{Klebanov,GinspargMoore,Polchinski})
has recently attracted a lot of attention \cite{c=1}
as the simplest string theory from which
we might learn something useful.

The $c=1$ matrix model is equivalent to a theory of free fermions $\Psi$
in 1+1 dimensional spacetime with the Hamiltonian
\footnote{
More precisely we should consider the field theory,
and the Hamiltonian is $\int dx \Psi^{\dag} H \Psi$,
where $\Psi(x)$ is the 2nd quantized fermion field.
We will use the notation of quantum mechanics
although everything can be extended to the 2nd quantized theory.
}
\be
H = -\frac{1}{2}\del_x^2 + V(x), \quad
V(x) = -\frac{x^2}{2}.
\ee
It is dual to 2 dimensional bosonic string theory or type 0B theory
\cite{TT,DKKMMS} depending on whether we fill one side or both sides
of the potential.
By filling the Fermi sea differently and an orbifolding,
the same model was also conjectured to be
dual to type \IIB theory \cite{GTT}.

A slight deformation of the potential
\be \label{H}
V(x) = -\frac{x^2}{2} + \frac{M}{2x^2}.
\ee
was conjectured \cite{JevickiYoneya}
to lead to bosonic string theory
in 2 dimensions in the background of a black hole of mass $M$.
More recently it was also conjectured \cite{TT,DKKMMS}
to be dual to type 0A string theory
in the linear dilaton background with RR electric field proportional to $q$,
where
\be \label{Mq}
M = q^2-\frac{1}{4}.
\ee

In the limit $M\rightarrow\infty$ (or $x\rightarrow 0$),
we can scale $x$ and ignore the quadratic term in the potential.
Hence
\be \label{AdS2V}
V(x) = \frac{M}{2x^2}.
\ee
This model was conjectured \cite{Strominger} to be dual to
type 0A string theory in the $AdS_2$ background.
The isometry group $SL(2, \mathbb{R})$ of $AdS_2$
is identified with a symmetry algebra of this matrix model.
The generators are
\be
H = \frac{1}{2}\left(p^2 + \frac{M}{x^2}\right), \quad
K = \frac{1}{2}x^2, \quad
D = -\frac{1}{4}\left(xp+px\right),
\label{HKD}
\ee
where $p = -i\del_x$ is the conjugate momentum of $x$.
They satisfy the $SL(2,\mathbb{R})$ Lie algebra
\be
[H, D]=iH, \quad [K, D]=-iK, \quad [H,K]=2iD.
\ee

The ground states of matrix models are specified by a single parameter $\mu$
which is the energy at the surface of the Fermi sea.
It is dual to the amplitude of a static tachyon background.
In the $AdS_2$ model the potential (\ref{AdS2V})
is bounded from below at zero, so $\mu \geq 0$.
The ground states with $\mu > 0$ spontaneously break
the $SL(2,\mathbb{R})$ symmetry.
Only the unique state with no fermion ($\mu = 0$) preserves the isometry,
and is matched to the invariant vacuum of $AdS_2$.

One might be puzzled by the fact that the $SL(2,\mathbb{R})$ symmetry
exists also for other matrix models as part of the $W_{\infty}$ algebra.
But of course $SL(2,\mathbb{R})$ can not be the isometry of
the dual target spaces which are asymptotically flat.
The resolution is that the $SL(2,\mathbb{R})$ symmetry
can be understood as the isometry of the matrix model
in a way we will explain later,
and the isometry is spontaneously broken by the ground state chosen
for the other matrix models.
Nevertheless, we will see that the $SL(2,\mathbb{R})$ symmetry
has important implications also for other matrix models.

Utilizing the $SL(2,\mathbb{R})$ generators,
the authors of \cite{AFF} showed that all quantum theories with
Hamiltonians of the form
\be \label{V0}
H = \frac{p^2}{2} + V(x), \quad V(x) = \frac{a}{2} x^2 + \frac{M}{2 x^2}
\ee
with the same parameter $M$, but arbitrary coefficient $a$,
are related to one another by coordinate transformations.
This has many striking implications.
For $M = 0$, the undeformed matrix model ($a = -1$) is in some sense
equivalent to the simple harmonic oscillator ($a = 1$),
and also to the theory without any potential ($a = 0$).
For $M > 0$, the deformed matrix model ($a = -1$) is in some way
the same as type 0A theory in $AdS_2$ ($a = 0$).
Yet these theories have completely different asymptotic behaviors!

As the potentials in (\ref{H}) and (\ref{AdS2V})
can be viewed as the same theory written in two sets of coordinates,
the $\mu\rightarrow -\infty$ limit of type 0A matrix model
can be identified with the vacuum of the $AdS_2$ matrix model.
In other words, type 0A theory in $AdS_2$ background
should be identified with the result of tachyon condensation
from the linear dilaton background.
\footnote{As the closed string tachyons are massless in 2 dimensions,
tachyon condensation might never happen.
What we mean here is to tune the parameter $\mu$
of the tachyon field by hand to $-\infty$.}
Due to the similarity between type 0A and 0B,
we conjecture that the same is true for type 0B theory as well.

The coordinate transformations relating theories
with different $a$ in (\ref{V0}) are not always bijective.
Some coordinate systems only cover a small part of the spacetime
defined by another set of coordinates.
We will see that properties of coordinate transformations of the matrix model
reflect those of the target space in string theory.
It was proposed \cite{Strominger} that
the Hamiltonian (\ref{V0}) with $a = 1$ (i.e., $H+K$)
is dual to $AdS_2$ in global coordinates,
while (\ref{AdS2V}) is dual to the Poincare patch,
which only covers half of the former.
Based on this proposal,
we will show that the RR flux background $q$ needs to be quantized
in order for the global time of $AdS_2$ to be compactified.

In addition, we will show that,
by adding a suitable (time-dependent) Fermi surface,
we can use the Hamiltonian (\ref{V0}) with any $a$ to
describe string theory in the linear dilaton background.
As the asymptotic behavior of particles for $a>0, a=0$ and $a<0$
are drastically different,
the phenomenology of each model appears to be very different,
although the encoded information is (almost) equivalent.
This reflects the observer dependence
familiar in the context of general relativity.
By analogy with $AdS_2$,
there should exist a coordinate transformation
(possibly combined with field redefinitions)
of string theory/supergravity in the linear dilaton background
which extends the spacetime beyond the region manifest in the old description.

In the appendix \ref{Isometry} we prove that (\ref{V0}) exhausts
all possibilities of nontrivial isometries for
1 dimensional quantum mechanics with the standard kinetic term.
The proof that all theories defined by (\ref{V0})
with the same $M$ are equivalent at the quantum level
are given in appendix \ref{appKilling}.

\section{$SL(2,\mathbb{R})$ Symmetry} \label{SL2R}

The quantum mechanics defined by (\ref{AdS2V})
was extensively studied \cite{AFF}
as an example of quantum mechanics with conformal symmetry.
Assuming that the kinetic energy is standard $\frac{1}{2}p^2$,
by simple dimensional analysis one can see that
(\ref{AdS2V}) is the only potential consistent with conformal symmetry.
(Conformal quantum mechanics with more than one variables
and general kinetic terms were discussed in \cite{MichelsonStrominger}.)
The $SL(2,\mathbb{R})$ symmetry (\ref{HKD})
was found in \cite{AFF} as the conformal symmetry.

In this section we will first review how
the $SL(2,\mathbb{R})$ symmetry is a symmetry of
conformal transformations of the time variable
together with a time-dependent scaling of the spatial coordinate
for the matrix model \cite{AFF}.
This could come as a bit of surprise for those who
are not familiar with the results of \cite{AFF},
since the spacetime coordinates in the matrix model
are nonlocally related to the coordinates of target space.
(The nonlocal transformation are determined by the leg factors
\cite{Polchinski,JevickiYoneya}.)

Then we will use the Lie algebra generators,
which are dual to Killing vectors in $AdS_2$,
to define new time coordinates,
and find that the new Hamiltonians are all of the form (\ref{V0})
with the same $M$.

\subsection{Isometry of Quantum Mechanics} \label{iqm}

An infinitesimal general coordinate transformation is generated
by a differential operator of the form
\be \label{D}
\D = iA\del_t + iB\del_x + C,
\ee
where $A, B, C$ are functions of $x$ and $t$.
Classically $C$ can be dropped,
and the infinitesimal coordinate transformation is
\be
\d t = \eps A, \quad \d x = \eps B.
\ee
The final result will justify why $C$ should be allowed
for quantum mechanics.
$\D$ generates a symmetry if
\be \label{EOMD}
\EOM \D = \D' \EOM
\ee
for some well defined operator $\D'$
so that a solution of the Schr\"{o}dinger equation
is still a solution after the transformation defined by $\D$, i.e.,
\be
\EOM \psi = 0 \quad \Rightarrow \quad \EOM e^{i\eps \D}\psi = 0.
\ee
Note that $\D'$ does not have to equal $\D$.
The condition (\ref{EOMD}) is equivalent to
\be
[\EOM, \hat{\D}] = 0, \quad \hat{\D} = AH - Bp + C, \label{Dhat}
\ee
where $\hat{\D}$ is $\D$ with $i\del_t$ replaced by the Hamiltonian $H$.
($\D$ and $\hat{\D}$ are equivalent when acting on
solutions of the Schr\"{o}dinger equation.)
The generators (\ref{HKD}) are to be
identified with $\hat{\D}$'s at $t = 0$ for the matrix model.

We will refer to the coordinate transformations which
preserve the Schr\"{o}dinger equation
as the quantum mechanics isometry.
It is natural to ask when a quantum theory admits nontrivial isometry.
Any time-independent Hamiltonian is an isometry generator of time translation.
We prove in appendix \ref{Isometry} for
with the standard kinetic energy
$\frac{1}{2} p^2$ the only time-independent Hamiltonian
with more than one isometry generators is of the form (\ref{V0}).
There are 3 isometry generators $H, K, D$ (\ref{HKD}) for a generic $M$.
If $M = 0$, there are 2 additional generators,
which can be identified with the creation and annihilation operators
of the simple harmonic oscillator when $a>0$.

In terms of the coordinates in the $AdS_2$ theory (\ref{AdS2V}),
the $SL(2,\mathbb{R})$ isometry
is represented as the projective transformation of time
and a scaling of space \cite{AFF}
\be
t \rightarrow \frac{\a t + \b}{\g t + \d}, \quad
x \rightarrow \frac{1}{\g t + \d}x.
\ee

\subsection{Killing Operators and Coordinate Transformations} \label{KO}

In analogy with Riemannian geometry,
the symmetry generators preserving Schr\"{o}dinger equations
are reminiscent of Killing vectors,
which preserve the metric,
and it is natural to use them to define new time coordinates.
Of course,
one can always perform arbitrary general coordinate transformations
or change of variables to rewrite a theory.
The special features of the time coordinates chosen by Killing operators
are that their conjugate Hamiltonians are time-independent
and of the standard form $\frac{1}{2}p^2 + V(x)$.

Take a generic $SL(2,\mathbb{R})$ generator
\be \label{G}
G = \a H + \b D + \g K,
\ee
and we would like to introduce a new coordinate $\t$
such that the old and new Schr\"{o}dinger equations are equivalent
(up to unitary transformation)
\be \label{Schrodinger}
i \frac{\del}{\del t}\Psi = H\Psi \Longleftrightarrow
i \frac{\del}{\del \t}\Psi' = G\Psi', \quad \Psi' = U\Psi.
\ee
It turns out that, remarkably,
$\tau$ is simply a function of $t$ \cite{AFF}
\be \label{tautransf}
d\tau = \frac{dt}{\a + \b t + \g t^2}.
\ee
For the range of $t$ satisfying
\be \label{condont}
f(t) = (\a + \b t + \g t^2) > 0,
\ee
$\t$ is a legitimate reparametrization of time.
Furthermore, after a simultaneous scaling of the spatial coordinate
\be \label{stransf}
\s = \frac{x}{(\a + \b t + \g t^2)^{1/2}},
\ee
$G$ is again of the form (\ref{V0}) \cite{AFF}
\be \label{HG}
G = -\frac{p_\s^2}{2} + V_G, \quad
V_G = -\frac{\Delta}{8} \s^2 + \frac{M}{2\s^2},
\ee
where
\be
\Delta = \b^2 - 4 \a\g.
\ee
The new coordinate system is as good as the old one
as long as (\ref{condont}) is satisfied.

The result above is derived in appendix \ref{appKilling}.
Here we verify it in the classical theory.
The action for the potential (\ref{AdS2V}) is
\be
S = \int dt \left(\frac{1}{2}\dot{x}^2 - \frac{M}{2 x^2}\right).
\ee
From (\ref{tautransf}) and (\ref{stransf}), it follows that
\be
\frac{dx}{dt} = f^{-1/2}(\del_{\t}\s + \frac{1}{2}(\del_t f) \s).
\ee
Plugging it in the action and integrating by parts, we find
\be \label{S}
S = \left[\frac{1}{2}(\del_t f)x^2\right]_{\t_0}^{\t_1}
+ \int d\t \left(\frac{1}{2}(\del_{\t}\s)^2
- \frac{d}{8}\s^2 - \frac{M}{2 \s^2} \right).
\ee
This is exactly the action for the Hamiltonian $G$ (\ref{HG}).
Due to the boundary term in (\ref{S}),
a unitary transformation is involved in the quantum version.

It is a very intriguing fact that
all Hamiltonians of the form (\ref{V0})
are different descriptions of essentially
the same theory related by simple coordinate transformations
(\ref{tautransf}) and (\ref{stransf}).

\subsection{Coordinate Patches: $S_-\subset S_0\subset S_+$} \label{CP}

A redefinition (scaling) of the coordinates
($\s \rightarrow \lam\s$, $\t \rightarrow \lam^2 \t$)
has the effect of scaling the coefficient of the $\s^2$ term by $\lam^4$.
So essentially we have 3 classes of $G$ that
are not related to each other in a trivial way.
They are examplified by $(\a = 1, \b = 0, \g = 1)$,
$(\a = 1, \b = 0, \g = 0)$ and $(\a = 1, \b = 0, \g = -1)$.
For notation,
we will equip variables with subscripts $+, 0, -$,
according to the values of $\g$,
and refer to the corresponding matrix models
by $S_+, S_0$, and $S_-$, respectively.
The Killing operators associated to $S_+, S_0, S_-$ are $(H+K), H, (H-K)$.
Their potentials are
\be \label{pot}
S_+:\;\; V_+ = \frac{1}{2}x_{+}^2 + \frac{M}{2x_{+}^2}, \quad
S_0:\;\; V_0 = \frac{M}{2x_0^2}, \quad
S_-:\;\; V_- = -\frac{1}{2}x_{-}^2 + \frac{M}{2x_{-}^2}.
\ee

The coordinate transformations are given by
(\ref{tautransf}) and (\ref{stransf})
\bea \label{coord-transf}
t_0 = \tan(t_+) = \tanh(t_-), \quad
x_0 = \sec(t_+) x_+ = \mbox{sech}(t_-) x_-.
\eea
It follows that the momenta are related by
\be \label{p-transf}
p_0 = \cos(t_+)p_+ + \sin(t_+) x_+ = \cosh(t_-)p_- - \sinh(t_-)x_-.
\ee
From these expressions we can see that
the origins of all time coordinates coincide.
At $t=0$, the spatial coordinates and
conjugate momenta also coincide.

For later use, we list here the solutions of the classical equations of motion
\bea
x_+ &=& \sqrt{A_+^2\cos(2(t_+ - T_+))+\sqrt{A_+^4 + M}}, \label{x+} \\
x_0 &=& \sqrt{2E_0 (t_0-T_0)^2 + \frac{M}{2E_0}}, \label{x0} \\
x_- &=& \sqrt{A_-^2\cosh(2(t_- - T_-))\pm\sqrt{A_-^4 - M}}, \label{x-}
\eea
where we assumed that $M>0$,
and so the particle stays on the positive side of the real line.
The parameters are related by
\bea \label{A+T+}
&A_+^2 = \sqrt{\left(E_0(T_0^2-1)+\frac{M}{4E_0}\right)^2+4E_0^2 T_0^2}, \quad
T_+ = -\tan^{-1}\left(\frac{2E_0 T_0}{E_0(T_0^2-1)+\frac{M}{4E_0}}\right), \\
&A_-^2 = \sqrt{\left(E_0(T_0^2+1)+\frac{M}{4E_0}\right)^2-4E_0^2 T_0^2}, \quad
T_- = -\tanh^{-1}\left(\frac{2E_0 T_0}{E_0(T_0^2+1)+\frac{M}{4E_0}}\right).
\eea
For $M = 0$, the solutions are
\bea
x_+ &=& A(\sin(t_+) - T_0 \cos(t_+)), \label{x+1} \\
x_0 &=& A(t_0 - T_0), \\
x_- &=& A(\sinh(t_-) - T_0 \cosh(t_-)). \label{x-1}
\eea
These solutions are related to each other through (\ref{coord-transf}).
But they are not the $M\rightarrow 0$ limit of (\ref{x+}-\ref{x-}),
since the $M = 0$ theory is not continuously connected to finite $M$ theories.

The coordinate transformations (\ref{coord-transf})
are not 1-1 mappings.
Roughly speaking, the coordinate systems $S_-, S_0, S_+$
forms a cascade of patches
\be \label{SSS}
S_- \subset S_0 \subset S_+.
\ee
The whole range $\mathbb{R}$ of time for a smaller patch is mapped to
a finite interval in a larger patch.
More precisely,
\bea
&\left(\begin{array}{l}
-\infty < t_- < \infty \\ |x_-| < \infty
\end{array}\right) \rightarrow
\left(\begin{array}{l}
-1 < t_0 < 1 \\ |x_0| < \infty 
\end{array}\right) \rightarrow
\left(\begin{array}{l}
-\frac{\pi}{4} < t_+ < \frac{\pi}{4} \\ |x_+| < \infty
\end{array}\right), \\
&\left(\begin{array}{l}
-\infty < t_0 < \infty \\ |x_0| < \infty
\end{array}\right) \rightarrow
\left(\begin{array}{l}
-\frac{\pi}{2} < t_+ < \frac{\pi}{2} \\ |x_+| < \infty
\end{array}\right). \label{ttt}
\eea
One can check that the condition (\ref{condont}) is satisfied
and $t_0, t_+$, and $t_-$ are legitimate time coordinates
within the ranges shown above.

So far we have treated $t_+$ as a parameter of $\mathbb{R}$.
Later we will see that $t_+$ should be compactified on a unit circle.
But the hierarchical relation (\ref{SSS}) remains the same.

\subsection{Matrix Model Isometry vs. Spacetime Isometry}

Each of these coordinate systems in the matrix model is matched to
a coordinate system in the dual theory of $AdS_2$
according to the associated Killing operators.
The Killing vectors of $AdS_2$ are
\be
H = i\del_t, \quad
D = i(t\del_t + \s\del_{\s}), \quad
K = i((t^2+\s^2)\del_t + 2t\s \del_s),
\ee
where we used the same notation to identify $SL(2, \mathbb{R})$ generators
in $AdS_2$ and the matrix model.
Here $(t, \s)$ are the coordinates of the Poincare patch.
The generator $(H+K)$ can be written as $i\del_{\t}$
where $\t$ is time in the global coordinates $(\t, \om)$
\be \label{global}
\t \pm \om = 2\tan^{-1}(t\pm \s).
\ee
Similarly, defining another set of coordinates
\be
\t' \pm \om' = 2\tanh^{-1}(t\pm \s),
\ee
one finds $H-K = i\del_{\t'}$.
This Killing vector is time-like only in a small part of $AdS_2$
covered by the Poincare patch.
We note that the relation (\ref{SSS}) is
mimicked by their $AdS_2$ cousins.
Despite the duality which guarantees some sort of matching,
this is nontrivial because the duality map is nonlocal.

\section{Target Space Geometry}

\subsection{$AdS_2$ as Near Horizon Geometry} \label{nearhorizon}

The supergravity solution of type 0A theory
with background RR electric field proportional to $q$ is \cite{BL}
\bea
ds^2 &=& (1 + \frac{q^2}{8} (\Phi-\Phi_0-\frac{1}{2})e^{2\Phi})
(-dt^2 + d\s^2), \\
\s(\Phi) &=& \frac{1}{\sqrt{2}}\int^{\Phi}
\frac{d\Phi'}{1+\frac{q^2}{8}(\Phi'-\Phi_0-\frac{1}{2})e^{2\Phi'}}, \\
\Phi_0 &=& - \log\frac{q}{4},
\eea
which is asymptotically flat in the $\s\rightarrow \infty$ limit.
If we scale the spatial coordinate around the ``near horizon'' region
$\s \rightarrow \infty$,
we arrive at the $AdS_2$ geometry
\be
ds^2 = \frac{1}{4\s^2}(-dt^2 + d\s^2), \quad
\Phi = \Phi_0 = -\log\frac{q}{4}.
\ee

Correspondingly, the deformed matrix quantum mechanics dual to
type 0A theory with RR flux $q$ is defined by (\ref{H}).
In terms of the generators defined in (\ref{HKD}),
the Hamiltonian is $H-K$.
If we scale $x$, $H$ and $K$ are scaled accordingly
\be
x \rightarrow \lam x, \quad H \rightarrow \lam^{-2} H,
\quad K \rightarrow \lam^2 K.
\ee
Obviously, the Hamiltonian approaches to $H$ as we zoom in $x$.
Essentially, this is simply saying that
if we zoom in around the region very close to $x = 0$,
we can eventually ignore the $x^2$ term in the potential.
Similarly, any Fermi surface with $\mu \leq 0$ will eventually
become out of sight as we zoom in around $x = 0$.
Hence the $AdS_2$ background corresponds to the vacuum
without fermion in the matrix model.

\subsection{Connections Among Matrix Models}

According to Sec. \ref{SL2R},
the $AdS_2$ matrix model (\ref{AdS2V}),
which is of the form of $S_0$,
can also be written as $S_+$ or $S_-$.
The distinctive feature of the $AdS_2$ theory is
that the ground state has no fermion.
When the Fermi sea is filled to a finite energy $\mu$ in $S_-$,
the matrix model is dual to string theory in the linear dilaton background
with a tachyon field proportional to $\mu$.
This implies that the back-reaction of the tachyon field changes
the target space geometry from an asymptotically flat space to $AdS_2$
in the limit of tachyon condensation $\mu \rightarrow -\infty$.
By tunning the tachyon amplitude, we can interpolate between
Minskowski space ($\mu=0$) and $AdS_2$ ($\mu=-\infty$).
This gives an explicit example of how large fluctuations
of the Fermi sea correspond to changes of the background geometry.

For bosonic or type 0B theory, there is no analogous scaling argument
on the supergravity side for $AdS_2$ as in Sec. \ref{nearhorizon}.
However, for the undeformed matrix quantum mechanics,
the vacuum with no fermion is still $SL(2,\mathbb{R})$ invariant.
Based on the similarity among type 0A, 0B and bosonic theories,
we propose that $AdS_2$ is also the spacetime geometry for type 0B theory
in the limit $\mu\rightarrow -\infty$.

If the correspondence between $S_+$ and the global $AdS_2$ theory is correct,
we would expect that a similar correspondence should persist
when a Fermi sea is introduced.
That is, there should exist an alternative description of
2 dimensional string theory with the linear dilaton background,
which contains the old story as a partial description of the full theory.
Since complicated field redefinition is involved in
matching the matrix model with the spacetime physics,
it is possible that the incompleteness of the usual description
is not simply referring to the spacetime,
but to the set of observables in string theory.
Currently we do not have a candidate for this theory,
we only know that it should be dual to $S_+$ with
a time-dependent Fermi sea.
(The Fermi sea in $S_+$ which corresponds to a ground state in $S_-$
will be described in Sec. \ref{FS}.)

\subsection{Topology of $AdS_2$ and Quantization of RR Charge} \label{TQ}

$AdS_2$ can be defined by its embedding in $2+1$ dimensions
\be \label{AdS}
X_0^2 + X_{-1}^2 - X_1^2 = 1.
\ee
The Poincare patch is related to Cartesian coordinates via
the coordinate transformation
\be
\s = (X_{-1} - X_1)^{-1}, \quad t = \s X_0.
\ee
It has the metric
\be
ds^2 = dX_0^2 + dX_{-1}^2 - dX_1^2 = \frac{1}{\s^2}(dt^2 - d\s^2).
\ee
The global coordinates of $AdS_2$ is defined by (\ref{global}),
for which the metric is
\be
ds^2 = \frac{1}{4\sin^2(\om)}(d\t^2 - d\om^2).
\ee
Translation in global time $\t$ is generated by
\be
L_0 = i\del_{\t} = H+K.
\ee
Hence $AdS_2$ in the Poincare patch is dual to $S_0$,
and in global coordinates to $S_+$ \cite{Strominger}.

According to the topology defined by (\ref{AdS}),
without extension to the covering space of $AdS_2$,
$\t$ is an angular variable,
and so
\be \label{top}
e^{i2\pi L_0} = 1,
\ee
which implies that the eigenvalues of $L_0$ have to be integers.
$L_0$'s eigenfunctions can be solved exactly \cite{Danielsson1}
and the eigenvalues of $H+K$ are
\be
E_n = n + |q|, \quad n = 0, 1, 2, \cdots,
\ee
where $q$ is the RR charge of type 0A theory.
The topological constraint (\ref{top}) then requires that
the RR charge $q$ is quantized.

With the spectrum given by integers,
the time variable $t_+$ of $S_+$ can be
naturally compactified on the unit circle
with $t_+\in(-\pi,\pi)$.
Recalling the image of $t_0$ in $S_+$ (\ref{ttt}),
we see that the spacetime of $S_0$ is half of the spacetime of $S_+$,
in perfect agreement with the relation between
the Poincare patch and global coordinates of $AdS_2$.

Since the matrix model for linear dilaton background
is equivalent to adding a certain Fermi sea background
in the $AdS_2$ matrix model, as we argued earlier,
the quantization of $q$ for $AdS_2$ ensures
the quantization of $q$ for other matrix models.

As a final remark on this issue, for $M > 0$,
the classical description of $S_+$ has a period of $\pi$ instead of $2\pi$
as shown by (\ref{x+}).
\footnote{
For $M = 0$,
the classical period of $t_+$ is already $2\pi$.
}
Hence the classical theory of $S_+$ is completely equivalent to that of $S_0$.
We need to examine the wave functions of $S_+$ in order to see that
the period of $t_+$ is actually $2\pi$.
On the other hand, from the viewpoint of dual theories,
it is a classical statement that the Poicare patch is half of the global $AdS_2$.
This is hence an example of the fact that sometimes
$AdS/CFT$ duality matches classical effects to quantum effects.

\section{Fermi Sea} \label{FS}

The ground state of a matrix model is a Fermi sea filled to an energy $\mu$.
In another theory this state is mapped to a time-varying Fermi sea,
which can not be viewed as a small fluctuation
over a static Fermi sea in view of the new Hamiltonian.
A configuration easy to discuss in one theory
may be complicated for another.

A generic Fermi surface can be described by its boundary
in the phase space $f(x, p, t) = 0$.
As every point on the boundary has to follow the trajectories
determined by the equation of motion
(\ref{x+}-\ref{x-}) and (\ref{x+1}-\ref{x-1}),
one can always rewrite $f(x, p, t)$ as a function of two variables.
The two variables can be any two constants of motion,
such as $(E_0, T_0)$ in (\ref{x0}).
Or instead we can use the coordinate and momentum at $t = 0$.
Whichever variables we choose,
we can use (\ref{coord-transf}) and (\ref{p-transf})
to switch the descriptions of a Fermi sea.
The benefit of using the phase space variables at $t = 0$
is that they are identical in all theories.

\begin{figure}
\begin{minipage}{70mm}
\begin{center}
\includegraphics[width=6cm]{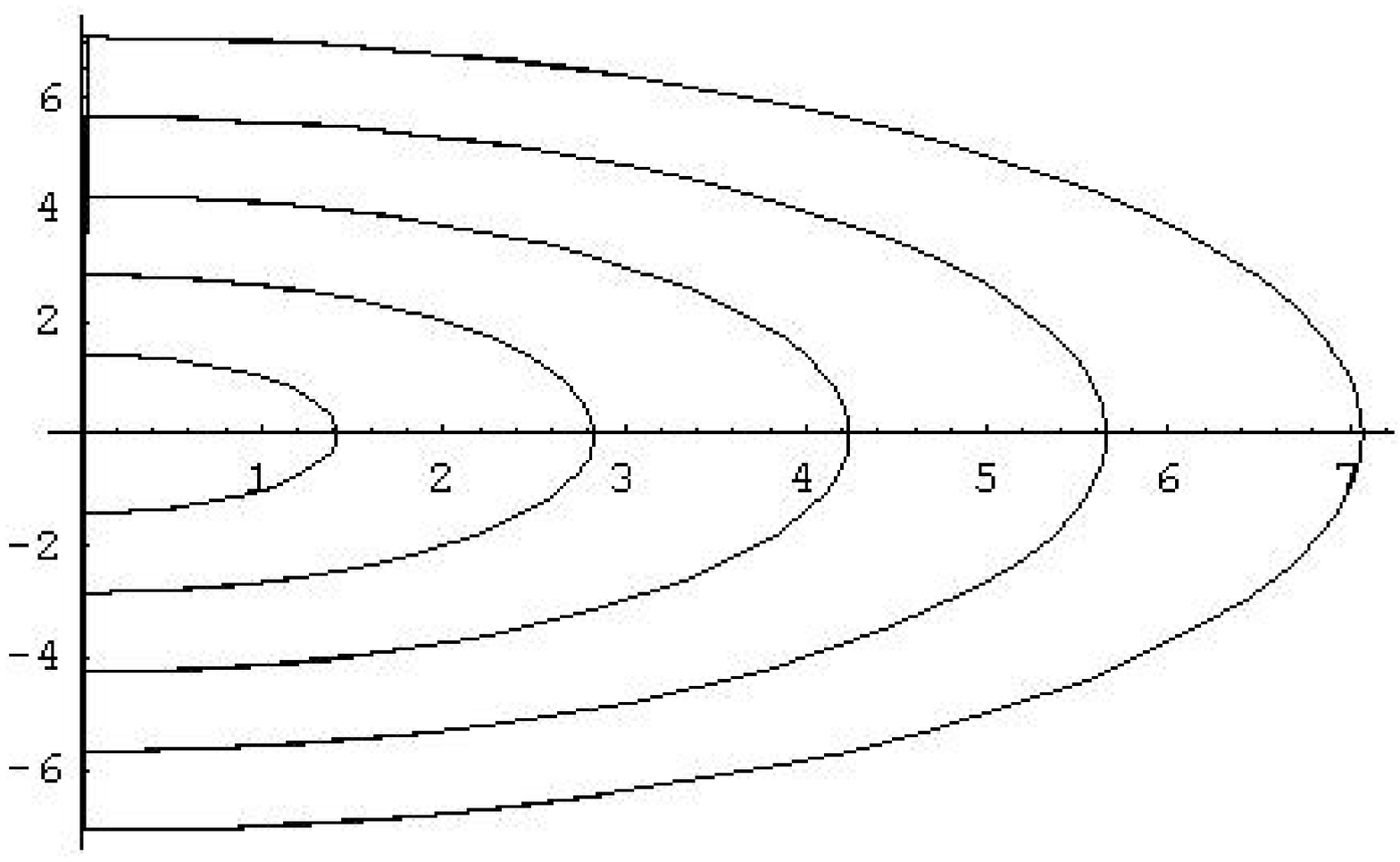}
\caption{ Phase space of $S_+$ for $M = 0$ } \label{PS+0}
\end{center}
\end{minipage}
\hspace*{10mm}
\begin{minipage}{70mm}
\begin{center}
\includegraphics[width=6cm]{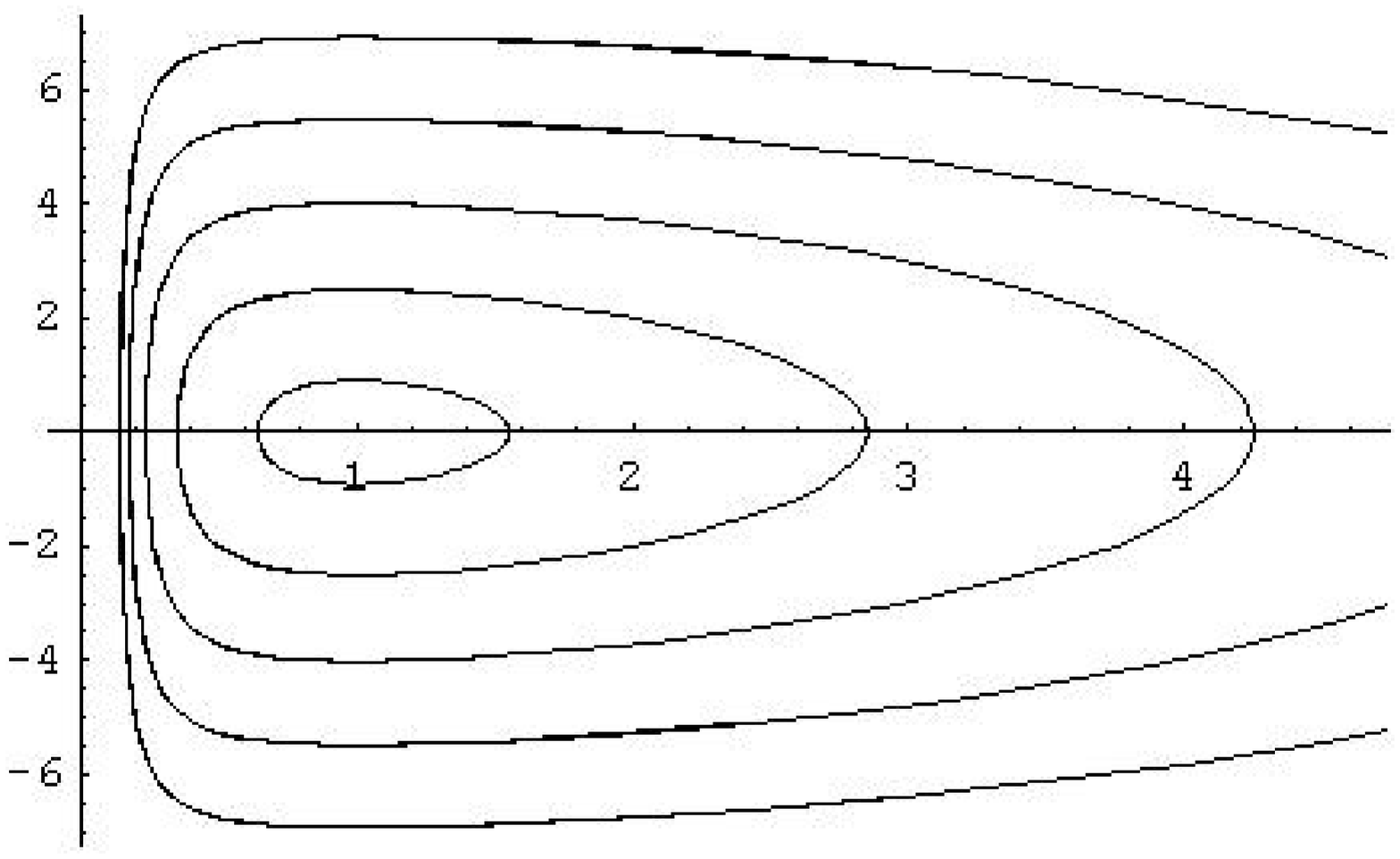}
\caption{ Phase space of $S_+$ for $M > 0$ } \label{PS+M}
\end{center}
\end{minipage}
\hspace*{10mm}
\end{figure}

For $M = 0$,
the phase space evolution for $S_+$ is simply
the circular periodic motion of the simple harmonic oscillator.
(See Fig.\ref{PS+0}.)
\footnote{
The lengths of curves are different because
they are particle trajectories for the same period of time,
i.e., faster particles leave a longer curve.
For $M = 0$ the phase spaces have left-right symmetry
and only the right half is plotted.
For $M > 0$ only $x > 0$ is allowed.
}
It is a periodic motion along deformed circles if $M > 0$.
(See Fig.\ref{PS+M}.)
Classically the period for $M = 0$ is twice that for $M > 0$.

\begin{figure}
\begin{minipage}{70mm}
\begin{center}
\includegraphics[width=6cm]{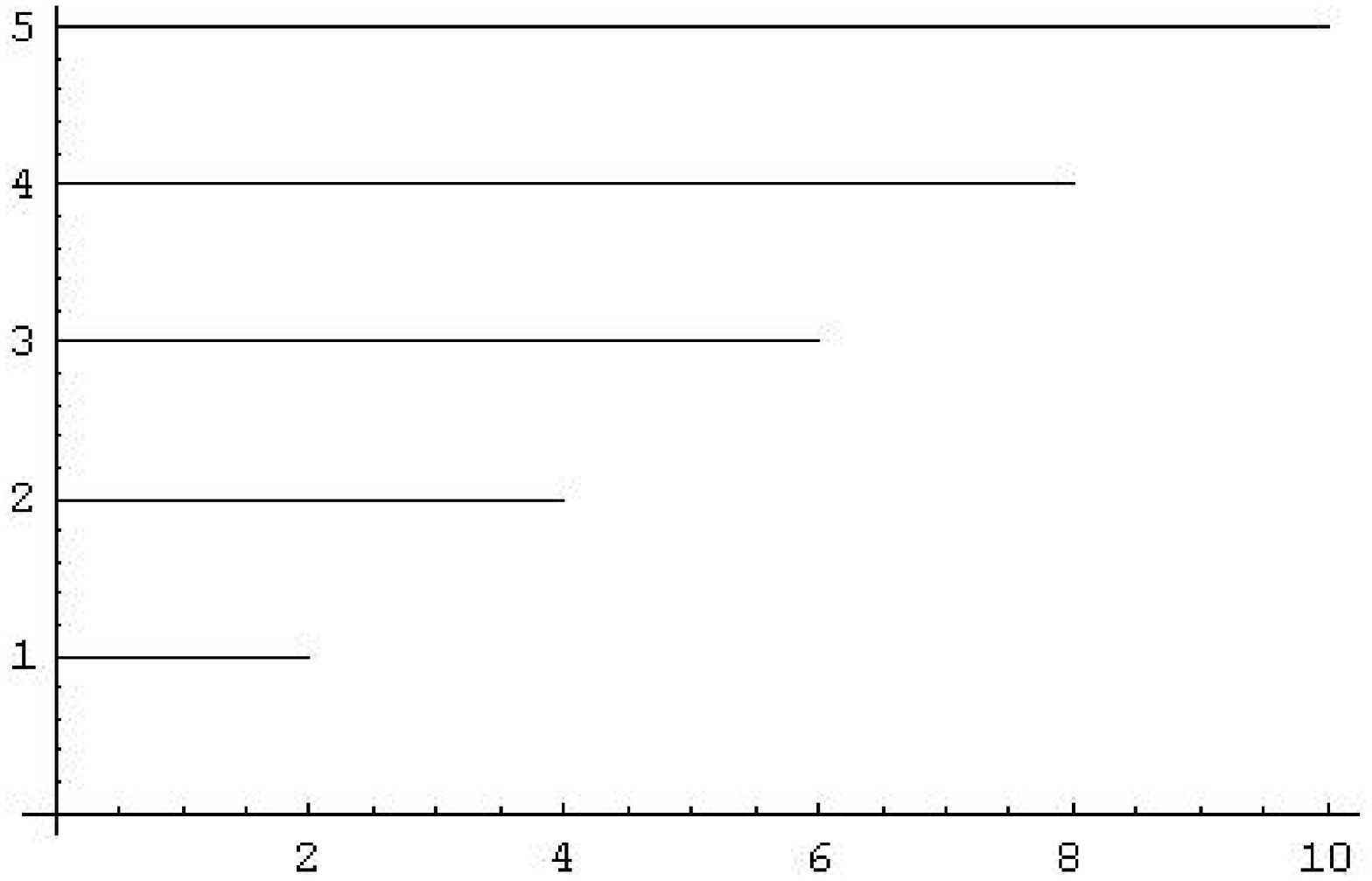}
\caption{ Phase space of $S_+$ for $M = 0$ } \label{PS00}
\end{center}
\end{minipage}
\hspace*{10mm}
\begin{minipage}{70mm}
\begin{center}
\includegraphics[width=6cm]{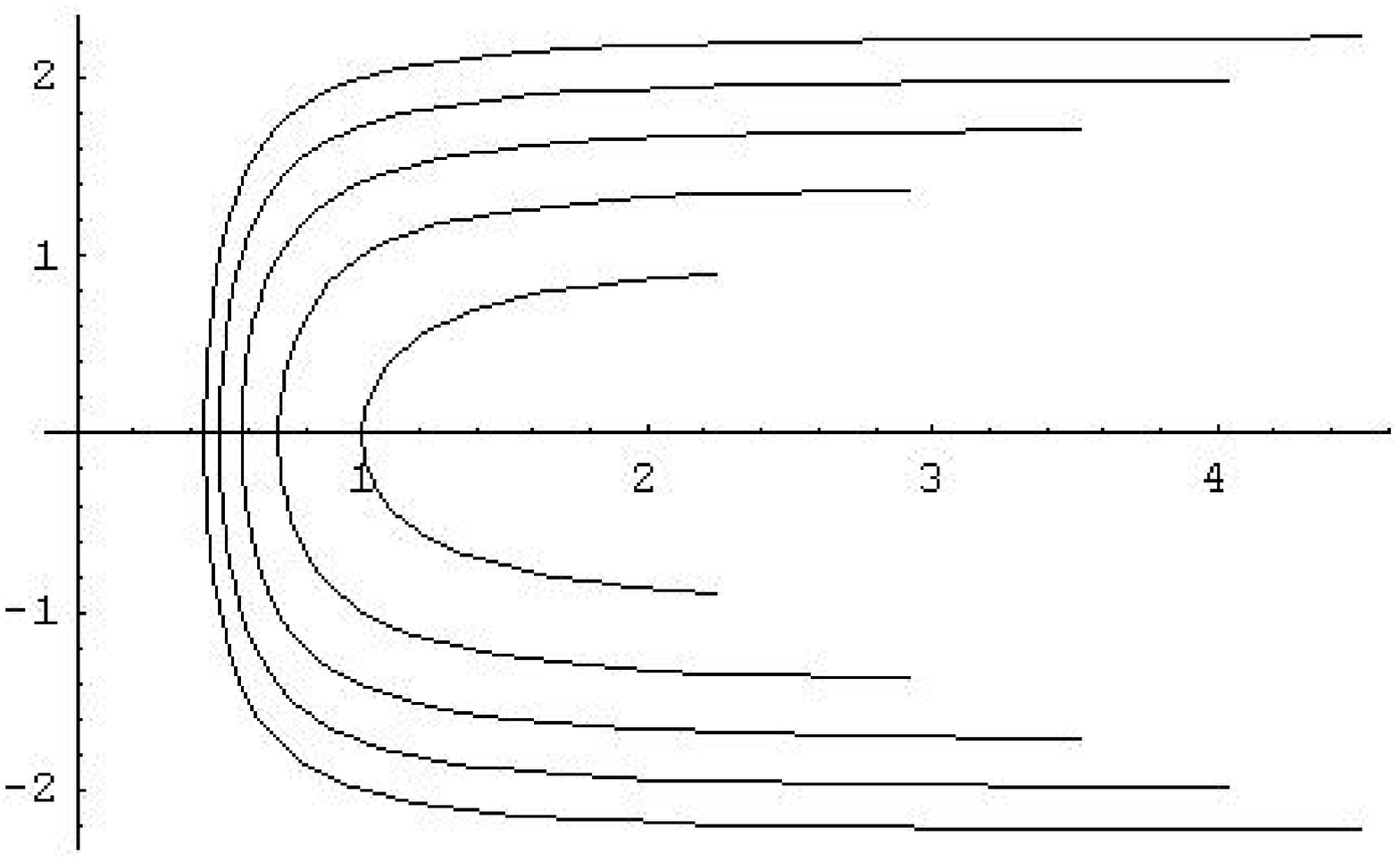}
\caption{ Phase space of $S_0$ for $M > 0$ } \label{PS0M}
\end{center}
\end{minipage}
\hspace*{10mm}
\end{figure}

The phase space evolution for $S_0$ is
straight horizontal motion along the $x_0$ axis if $M = 0$.
(See Fig.\ref{PS00}.)
For $M > 0$ a point in the phase space turns around at
the infinite potential wall.
The trajectories are asymptotically horizontal lines.
(See Fig.\ref{PS0M}.)

\begin{figure}
\begin{minipage}{70mm}
\begin{center}
\includegraphics[width=6cm]{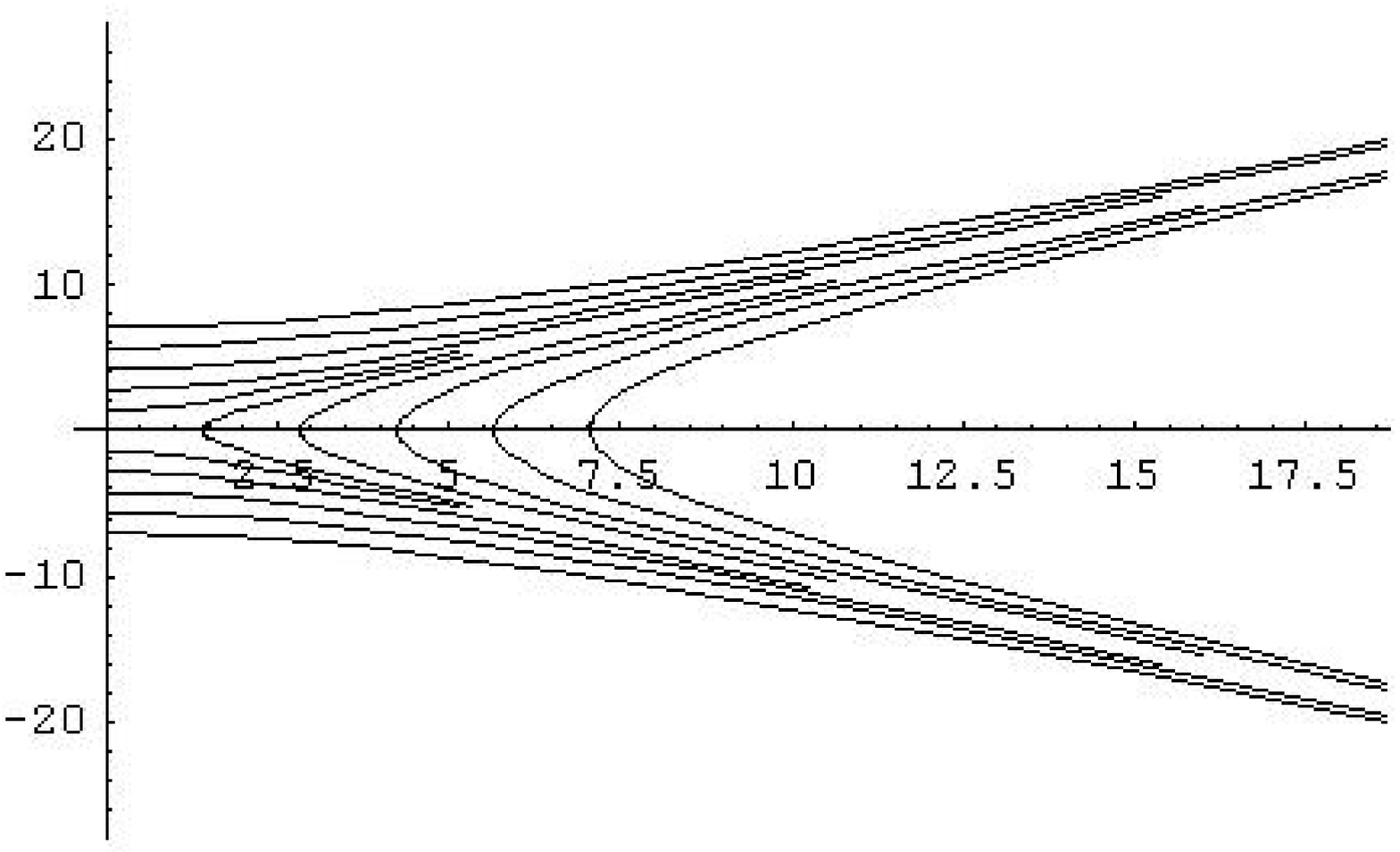}
\caption{ Phase space of $S_+$ for $M = 0$ } \label{PS-0}
\end{center}
\end{minipage}
\hspace*{10mm}
\begin{minipage}{70mm}
\begin{center}
\includegraphics[width=6cm]{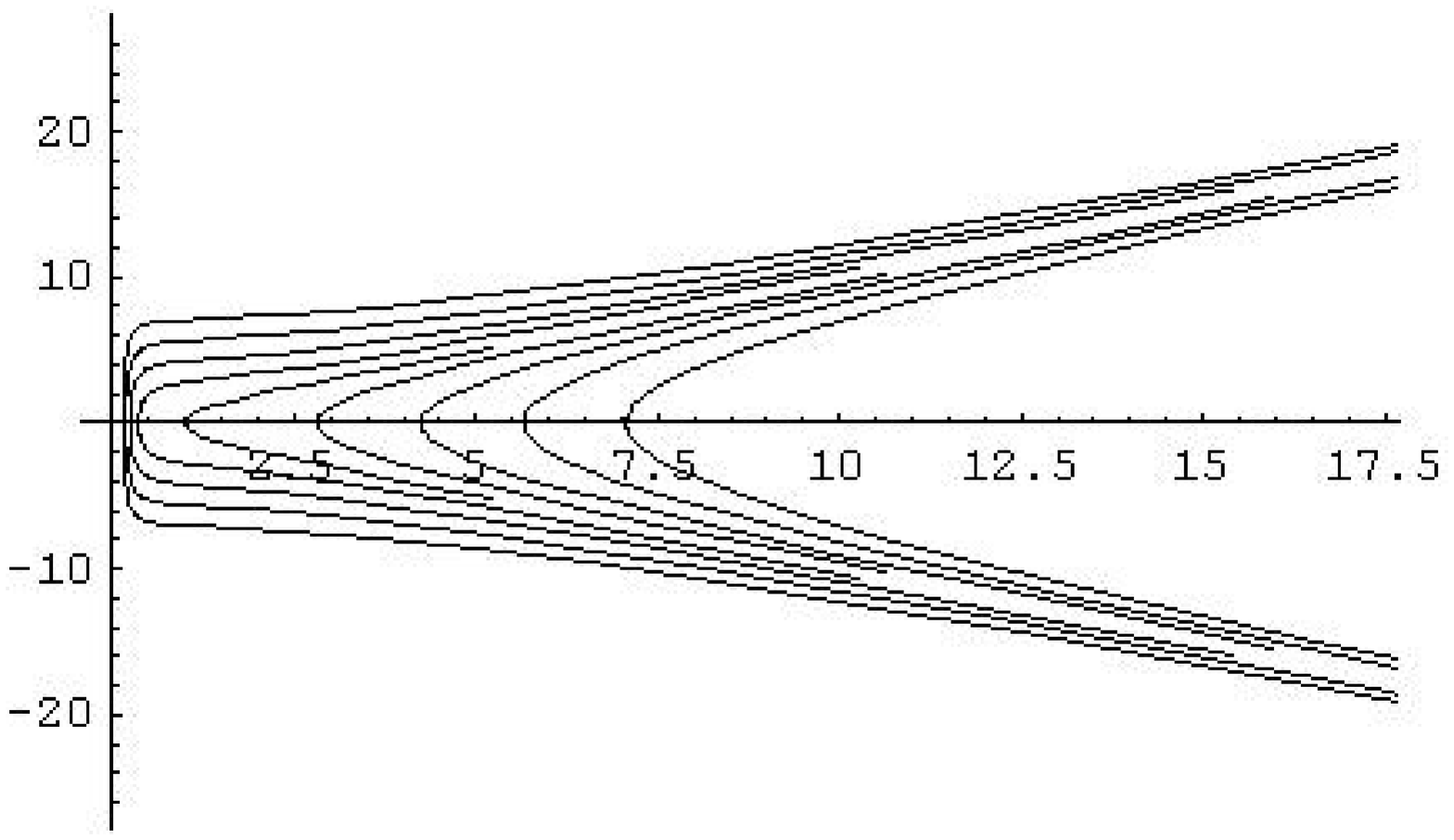}
\caption{ Phase space of $S_-$ for $M > 0$ } \label{PS-M}
\end{center}
\end{minipage}
\hspace*{10mm}
\end{figure}

For $S_-$, a point in the phase space always moves along
a hyperbolic curve.
(See Fig.\ref{PS-0} and Fig.\ref{PS-M}.)

It is then easy to visualize the Fermi surface
in a different coordinate system.
The phase space configuration of a Fermi surface
at $t=0$ in one coordinate system
is identical in another coordinate system.
As long as we know how each point in the phase space evolves with time,
we can easily figure out the evolution of the Fermi surface in any theory.

For example, the static Fermi sea in the matrix model
for bosonic strings is bounded by a hyperbolic curve in the phase space.
It is static for $S_-$, and is rotating around the origin for $S_+$.
In general a static Fermi sea is time-dependent in another coordinate system.
The origin of the phase space for $M = 0$ is the only case
that is static in all theories.

In \cite{KS}, interesting time-dependent solutions of the Fermi sea
were found in the matrix model.
Some of them describe tachyon condensation.
For instance, the solution for light-like tachyon condensation,
\be
(x + p + 2 \lam e^t)(x - p) = g_s^{-1}
\ee
describes a Fermi sea which is at rest
and filled up to the energy $g_s^{-1}$ in the infinite past,
but has all fermions removed to the infinities in the infinite future.
However, in the $S_+$ theory all solutions are periodic.
If tachyon condensation happens during half the cycle,
the reverse process must take over the other half of the cycle.
This is consistent with the fact that
tachyons in 2 dimensions are not tachyonic.


\section{Discussions}

In this paper we clarified the meaning of the $SL(2,\mathbb{R})$ symmetry
as an isometry of the matrix model.
We also used its Lie algebra generators,
interpreted as Killing vectors of $AdS_2$ in the dual theory,
to define new coordinate systems in which
the matrix model takes different Hamiltonians.
This reflects the different appearances of $AdS_2$
in different coordinate patches.
It turns out that all matrix models (\ref{V0}) are in some sense equivalent,
and a choice of the string theory background corresponds
to a choice of the Fermi sea.
The moduli space of string theory is mapped to
the space of all Fermi sea configurations
(including the time-dependent ones).

As a supportive evidence of our interpretation,
we used $AdS_2$'s topological property to quantize
the D0-brane charge $q$ for the type 0A matrix model.
We also find that the relation between
the Poincare patch and global coordinates of $AdS_2$
is faithfully inherited by the map between time coordinates
of corresponding matrix models.

Finally, let us examine more carefully to what extent
we can claim the quantum equivalence among $S_+, S_0, S_-$.
First, the Hilbert spaces, if defined as the spaces of
normalizable wave functions,
are identical even at the quantum level for all finite $t$
(whenever the coordinate transformation is not singular),
although the energy eigenfunctions are of course different.
Yet if we further restrict ourselves to states with finite energies,
the Hilbert spaces are different among $S_+, S_0, S_-$.
For instance, according to (\ref{x+}) and (\ref{A+T+}),
\be
E_+ = E_0(T_0 + 1)^2 + \frac{M}{4E_0},
\ee
so $E_+$ diverges whenever $E_0 = 0$ or $T_0\rightarrow \pm\infty$,
which are finite energies in $S_0$.
This is what one should expect as a generic phenomenon
associated with spacetime coordinate transformations.

On the other hand, we see from our discussion in Sec.\ref{CP}
that it can not be a complete equivalence,
since the coordinate transformations are not bijective.
In fact we already have an example of inequivalence:
the quantization of RR flux due to compactification of $t_+$ in $S_+$
can not be repeated in $S_0$ or $S_-$
since the latter have no access to the full range of $t_+$.
Similarly, we expect that $S_+$ will be superior to $S_0$ and $S_-$
when we consider certain orbifolds of $AdS_2$,
and that this conclusion can be carried over to the linear dilaton background
when the appropriate Fermi sea is introduced.

\section*{Acknowledgment}

The author thanks Allan Adams, Joanna Karczmarek, Lubos Motl,
Andy Strominger, Tadashi Takayanagi and John Wang
for helpful discussions.
This work is supported in part by
the National Science Council, Taiwan, R.O.C.
the Center for Theoretical Physics
at National Taiwan University,
and the CosPA project of the Ministry of Education, Taiwan, R.O.C.

\appendix

\section{Solutions to the Isometry Condition} \label{Isometry}

The isometry condition (\ref{EOMD})
can be solved for any given Hamiltonian $H$.
In fact it is possible to use the isometry condition
to find all Hamiltonians with nontrivial isometries.

Following Sec. \ref{iqm},
if $H$ has a quadratic term in $\del_x$,
we can compare the coefficients of $\del_t$ on both sides of (\ref{EOMD}).
It implies that
\be
\quad \D' = \D + i\frac{\del}{\del t} A.
\ee
Then we compare the coefficient of $\del_t\del_x$,
and see that we must have $\del_x A = 0$,
that is, $A$ is a function of $t$ only.
Hamiltonians with only linear terms in $\del_x$
will not be considered in this paper.
But we expect that similar approach will apply.

Matching the coefficients of $\del_x^n$ for each $n$
in the isometry condition (\ref{EOMD}) gives
\bea
&\del_x B = \frac{1}{2}\del_t A, \quad \del_x C = \del_t B, \\
&i\del_t(C+AV) + iB\del_x V + \frac{1}{2}\del_x^2 C = 0. \label{eq3}
\eea
The first two can be easily solved
\be \label{BC}
B = \frac{1}{2}\del_t A(t) x + b(t), \quad
C = \frac{1}{4}\del_t^2 A(t) x^2 + \del_t b(t) x + c_0(t),
\ee
and after plugging these in (\ref{eq3}) we get
\be \label{3rdid}
\del_t A (V+\frac{1}{2}x\del_x V)+ b\del_x V +
\frac{1}{4}\del_t^3 A x^2 + \del_t^2 b x +
\del_t c_0 - \frac{i}{4}\del_t^2 A = 0.
\ee
Now we should consider separately the cases $\del_t A \neq 0$ and $\del_t A = 0$.

If $\del_t A \neq 0$,
we can shift $x$ by $x\rightarrow x - \frac{2b}{\del_t A}$
to absord $b$.
Hence we can assume $b(t) = 0$
without loss of generality.
We will still use $x$ to stand for the shifted coordinate.
The equations (\ref{BC}-\ref{3rdid}) are simplified to
\bea
&B = \frac{1}{2}\del_t A(t) x, \quad
C = \frac{1}{4}\del_t^2 A(t) x^2 + c_0(t), \label{BC1} \\
&\del_t A (V+\frac{1}{2}x\del_x V) +
\frac{1}{4}\del_t^3 A x^2 +
\del_t c_0 - \frac{i}{4}\del_t^2 A = 0. \label{3rdid1}
\eea
Eq. (\ref{3rdid1}) is a statement about the linear dependence
of $(V+\frac{x}{2}\del_x V), x^2, x, 1$ as functions of $x$.
Yet since the coefficients of them are functions of $t$,
the only chance for it to hold for all $t$ and $x$ is that
all coefficients are the same function of $t$ up to overall constant factors.
That is
\be \label{eqAc0}
\frac{\del_t^3 A}{\del_t A} = -8 v_2, \quad
\frac{-\del_t c_0 + \frac{i}{4}\del_t^2 A}{\del_t A} = v_0,
\ee
for some constants $v_2, v_0$.
The potential $V$ can then be easily solved from (\ref{3rdid1})
\be \label{V}
V = v_0 + v_2 x^2 + \frac{M}{2 x^2}.
\ee

The isometry generators $\D$ are defined by $A$, $B$ and $C$,
which are determined by (\ref{BC1}) and (\ref{eqAc0}).
For the potential (\ref{V}) with $v_2\neq 0$,
the result is that $\D$ is in general a linear combination of
the following 3 generators
\bea
&{\cal H} = i \del_t - v_0, \label{H+-} \\
&{\cal L}_+ = e^{\sqrt{-8v_2}t}\left( i\del_t +
\frac{i}{2}\sqrt{-8v_2}x\del_x - 2v_2 x^2
-(v_0 - \frac{i}{4}\sqrt{-8v_2}) \right), \label{L+} \\
&{\cal L}_- = e^{-\sqrt{-8v_2}t}\left( i\del_t -
\frac{i}{2}\sqrt{-8v_2}x\del_x - 2v_2 x^2
-(v_0 + \frac{i}{4}\sqrt{-8v_2}) \right). \label{L-}
\eea
They satisfy the $SL(2,\mathbb{R})$ Lie algebra
\be
[{\cal H}, {\cal L}_{\pm}] = \pm i\sqrt{-8v_2}{\cal L}_{\pm},
\quad
[{\cal L}_+, {\cal L}_-] = -2i\sqrt{-8v_2}{\cal H}.
\ee
For the potential (\ref{V}) with $v_2 = 0$, i.e.,
\be \label{v2=0}
V = v_0 + \frac{M}{2 x^2},
\ee
they are
\bea
&{\cal H} = i \del_t - v_0, \label{H2} \\
&{\cal L}_1 = it\del_t + \frac{i}{2}x\del_x - v_0 t + \frac{i}{4},
\label{L1} \\
&{\cal L}_2 = \frac{i}{2}t^2\del_t + \frac{i}{2}tx\del_x
+ \frac{x^2}{4} - \frac{v_0}{2} t^2 + \frac{i}{4} t.
\label{L2}
\eea
They realize a different set of generators of the same algebra
\be \label{HL1L2}
[{\cal H}, {\cal L}_1 ] = i{\cal H}, \quad
[{\cal H}, {\cal L}_2 ] = i{\cal L}_1, \quad
[{\cal L}_1, {\cal L}_2 ] = i{\cal L}_2.
\ee

If on the other hand $\del_t A = 0$, then $A$ is a constant,
and $b$ has to be nonzero for the existence
of any additional isometry generator in addition to ${\cal H}$.
Due to $\del_t A = 0$, (\ref{3rdid}) becomes
\be
b\del_x V + \del_t^2 b x + \del_t c = 0.
\ee
Again this implies that $b, \del_t^2 b$ and $\del_t c$
can only differ by constant factors.
Thus $V$ has to be of the form
\be
V = v_2 x^2 + v_0,
\ee
and the symmetry generators are
\be \label{M}
{\cal M}_{\pm} = e^{\pm\sqrt{-2v_2}t}\left(
i\del_x \pm \sqrt{-2v_2} x \right),
\ee
in addition to ${\cal H} = i\del_t$.
${\cal M}_{\pm}$ are space-like Killing operators.
They satisfy the algebra
\be \label{crann}
[{\cal H}, {\cal M}_{\pm}] = \pm i \sqrt{-2v_2} {\cal M}_{\pm}, \quad
[{\cal M}_+, {\cal M}_-] = -2i\sqrt{-2v_2}.
\ee
The generators ${\cal L}_{\pm}$ (\ref{L+}, \ref{L-})
can still be defined when $v_{-2} = 0$.
Their commutation relations with ${\cal M}_{\pm}$ are
\be
[{\cal L}_{\pm}, {\cal M}_{\pm}] = 0, \quad
[{\cal L}_{\pm}, {\cal M}_{\mp}] = \mp 2i \sqrt{-2v_2} {\cal M}_{\pm}.
\ee

To summarize, for $M = 0$ in (\ref{V0}),
there are a total of 5 isometry generators
${\cal H}, {\cal L}_{\pm}$ and ${\cal M}_{\pm}$.
The two new generators ${\cal M}_{\pm}$
are in fact simply the creation and annihilation operators
of the simple harmonic oscillator when $v_2 > 0$.
(When $M \neq 0$, the $SL(2,\mathbb{R})$ algebra
is also very useful for studying the spectrum \cite{AFF,Danielsson1}.)
When $v_2 < 0$, they can be used to construct
the discrete spectrum of imaginary energies in the matrix model.

If $v_2 = 0$, the new generators (\ref{M}) reduce to
\be
{\cal M}_1 = i\del_x, \quad {\cal M}_2 = it\del_x + x.
\ee
The algebra is defined by (\ref{HL1L2}) and
\bea
&[{\cal H}, {\cal M}_1] = 0, \quad
[{\cal H}, {\cal M}_2] = i{\cal M}_1, \quad
[{\cal M}_1, {\cal M}_2] = i, \\
&[{\cal L}_1, {\cal M}_1] = -\frac{i}{2}{\cal M}_1, \quad
[{\cal L}_1, {\cal M}_2] = \frac{i}{2}{\cal M}_2, \\
&[{\cal L}_2, {\cal M}_1] = -\frac{i}{2}{\cal M}_2, \quad
[{\cal L}_2, {\cal M}_2] = 0.
\eea

Using the fact that all Hamiltonians with potentials
of the form (\ref{V}) with the same $M$
are related to each other by coordinate transformations \cite{AFF}
(see next section),
the generators (\ref{H+-}), (\ref{L+}), (\ref{L-})
are related to the generators (\ref{H2}), (\ref{L1}), (\ref{L2})
by coordinate transformations.
By unitary transformations of $e^{iv_0 t}$,
we can always set $v_0 = 0$.
Then we see that the generators in (\ref{H2}-\ref{L2})
are related to $H, K, D$ in (\ref{HKD}) as
\be
({\cal H}, {\cal L}_1, 2{\cal L}_2) \rightarrow (H, D, K),
\ee
respectively,
by setting $t = 0$ and replacing $i\del_t$ by $H$.

\section{Killing Operator as Hamiltonian} \label{appKilling}

By suitably choosing a new time coordinate $\t$,
we can rewrite the isometry generator $\D$
\be \label{D1}
\D = \a{\cal H} + \b{\cal L}_1 + 2\g{\cal L}_2 =
if(t) \del_t + \frac{1}{2}(\del_t f(t)) x\del_x
+ \frac{i}{4}(\del_t f(t) - i2\g x^2)
\ee
as
\be \label{D2}
i\del_{\t},
\ee
where
\be
f(t) = \a + \b t + \g t^2
\ee
for some constants $\a, \b, \g$,
up to unitary transformations,
so that $\t$ is the coordinate translated by $\D$.
Apparently, for the new Schr\"{o}dinger equation
\be
(i\del_\t - H_{\t}) \psi = 0
\ee
to be equivalent to the old one,
$H_{\t}$ should be equivalent to $\hat{\D}$,
which is $\D$ with $i\del_t$ replaced by $H$.

Comparing (\ref{D1}) with (\ref{D2}), we demand that
\be
\frac{\del t}{\del \t} = f(t), \quad
\frac{\del x}{\del \t} = \frac{1}{2}\del_t f.
\ee
They are solved by (\ref{tautransf}) and (\ref{stransf}),
or more explicitly
\be
\tau = \tau(t) =
\frac{2}{\sqrt{\Delta}} \tan^{-1}\left(\frac{f'(t)}{\sqrt{\Delta}}\right),
\quad \s = f^{-1/2} x, \label{sy},
\ee
where $\Delta = 4\a\g-\b^2$.
Here $\s$ is the new spatial coordinate.
Obviously there is a freedom to shift $\tau$ by a constant.

For (\ref{sy}) to make sense we need to assume $\g\neq 0$.
But they are still valid even if $\Delta < 0$,
using $\tan(i\th) = i\tanh(\th)$.
In terms of the new variables, the equation of motion becomes
\be
\EOM = f^{-1}\left(
i\del_{\t}+\frac{i}{4}(\del_t f)+
\frac{1}{2}(\del_{\s}-\frac{i}{2}(\del_t f)\s)^2
+\frac{1}{8}(\del_t f)^2\s^2-\frac{v_{-2}}{\s^2}
\right) = f^{-1}\EOM'.
\ee
One can find the new Hamiltonian by reading it off from
the new equation of motion.
This expression can be simplified by a unitary transformation
\be
\psi = U\hat{\psi}, \quad U = e^{\frac{i}{4}(\del_t f)\s^2},
\ee
and a rescaling of the wave function to absorb
the time dependent measure
\be
\hat{\psi} = f^{1/4}\tilde{\psi}, \quad
\frac{dx}{d\s} = f^{-1/2}.
\ee
Let the new Hamiltonian be defined through
\be
i\del_{\t} - H_{\t} = f^{1/4}U^{\dag}\left(f\EOM\right)Uf^{-1/4},
\ee
We find
\be \label{Htau}
H_{\t} = -\frac{1}{2}\del_{\s}^2
+\frac{M}{2 \s^2}+\frac{\Delta}{8}\s^2.
\ee
Finally, one can check that
\be
U^{\dag}\D U = i\del_{\t},
\ee
as we aimed at in the beginning.
Compared with Sec. \ref{KO},
$H_{\t}$ (\ref{Htau}) was denoted $G$ in (\ref{HG}).

In the derivation above we see that,
in addition to the coordinate transformation,
a unitary transformation by $U$ and a scaling by $f^{1/4}$
is needed for the wave function to satisfy
Schr\"{o}dinger equation in the new coordinate system.


\vskip .8cm
\baselineskip 22pt
\begin{thebibliography}{99}
\itemsep 0pt

\bibitem{MM}
D.~J.~Gross and N.~Miljkovic,
``A Nonperturbative Solution Of D = 1 String Theory,''
Phys.\ Lett.\ B {\bf 238}, 217 (1990);
E.~Brezin, V.~A.~Kazakov and A.~B.~Zamolodchikov,
 ``Scaling Violation In A Field Theory Of Closed Strings In One Physical Dimension,''
Nucl.\ Phys.\ B {\bf 338}, 673 (1990);
P.~Ginsparg and J.~Zinn-Justin,
``2-D Gravity + 1-D Matter,''
Phys.\ Lett.\ B {\bf 240}, 333 (1990).

\bibitem{Klebanov}
I.~R.~Klebanov,
``String theory in two-dimensions,''
[hep-th/9108019].

\bibitem{GinspargMoore}
P.~Ginsparg and G.~W.~Moore,
``Lectures On 2-D Gravity And 2-D String Theory,''
[hep-th/9304011].

\bibitem{Polchinski}
J.~Polchinski,
``What is string theory?,''
[hep-th/9411028].

\bibitem{c=1}
J.~McGreevy and H.~Verlinde,
``Strings from tachyons: The c = 1 matrix reloaded,''
JHEP {\bf 0312}, 054 (2003)
[hep-th/0304224];
and for a long list of recent reference, see e.g. \cite{GTT}.

\bibitem{TT}
T.~Takayanagi and N.~Toumbas,
``A matrix model dual of type 0B string theory in two dimensions,''
JHEP {\bf 0307}, 064 (2003)
[hep-th/0307083].

\bibitem{DKKMMS}
M.~R.~Douglas, I.~R.~Klebanov, D.~Kutasov,
J.~Maldacena, E.~Martinec and N.~Seiberg,
``A new hat for the c = 1 matrix model,''
[hep-th/0307195].

\bibitem{GTT}
S.~Gukov, T.~Takayanagi and N.~Toumbas,
``Flux backgrounds in 2D string theory,''
[hep-th/0312208].

\bibitem{JevickiYoneya}
A.~Jevicki and T.~Yoneya,
``A Deformed matrix model and the black hole background
in two-dimensional string theory,''
Nucl.\ Phys.\ B {\bf 411}, 64 (1994)
[hep-th/9305109].

\bibitem{Strominger}
A.~Strominger,
``A matrix model for AdS(2),''
[hep-th/0312194].

\bibitem{AFF}
V.~de Alfaro, S.~Fubini and G.~Furlan,
``Conformal Invariance In Quantum Mechanics,''
Nuovo Cim.\ A {\bf 34}, 569 (1976).

\bibitem{MichelsonStrominger}
J.~Michelson and A.~Strominger,
``The geometry of (super)conformal quantum mechanics,''
Commun.\ Math.\ Phys.\  {\bf 213}, 1 (2000)
[hep-th/9907191].

\bibitem{BL}
N.~Berkovits, S.~Gukov and B.~C.~Vallilo,
``Superstrings in 2D backgrounds with R-R flux and new extremal black holes,''
Nucl.\ Phys.\ B {\bf 614}, 195 (2001)
[hep-th/0107140].

\bibitem{Danielsson1}
U.~H.~Danielsson,
``A Matrix model black hole,''
Nucl.\ Phys.\ B {\bf 410}, 395 (1993)
[hep-th/9306063].

\bibitem{KS}
J.~L.~Karczmarek and A.~Strominger,
``Matrix cosmology,''
[hep-th/0309138].

\end{thebibliography}
\end{document}